\documentclass[11pt]{article}

\usepackage{amssymb,graphicx,fullpage}

\begin{document}

\title{Stability and bifurcations  in an epidemic model with varying immunity period}

\author{K.B. Blyuss\thanks{Corresponding author. Email: k.blyuss@sussex.ac.uk}\hspace{0.2cm} and Y.N. Kyrychko
\\\\Department of Mathematics, University of Sussex,\\Brighton, BN1 9QH, United Kingdom}

\maketitle

\begin{abstract}
An epidemic model with distributed time delay is derived to describe the dynamics of infectious diseases with varying immunity. It is shown that solutions are always positive, and the model has at most two steady states: disease-free and endemic. It is proved that the disease-free equilibrium is locally and globally asymptotically stable. When an endemic equilibrium exists, it is possible to analytically prove its local and global stability using Lyapunov functionals. Bifurcation analysis is performed using DDE-BIFTOOL and traceDDE to investigate different dynamical regimes in the model using numerical continuation for different values of system parameters and different integral kernels.
\end{abstract}

\section{Introduction}

Recent years have witnessed a rapid increase in the use of mathematical models for better understanding of epidemics and disease dynamics. Mathematical models take into account main factors that govern development of a disease, such as transmission and recovery rates, and predict how the disease will spread over a period of time. Traditional epidemiological models divide the whole population into three classes of susceptible, infected and recovered individuals, and the spread of an epidemic is governed by the principle of mass action \cite{AM}. For certain diseases, such as, for example, sexually transmitted infections, it is important to account for the individuals who have been exposed to the infection but have not yet become infected, thus the whole population is split into four groups, including a separate group of exposed. The incidence rate with which individuals become infected is normally taken to be bilinear with respect to the susceptible and infected populations. However, there is some evidence that a bilinear incidence rate might not be an effective assumption for highly contagious diseases, where a high percentage of the whole population is infected, and the transition from susceptible to infected has to be represented by a non-linear function \cite{DvD,HvD,KM}.  

It is well known that the spread of many infectious diseases can be prevented by vaccination of the susceptible population. Furthermore, some infections provide recovered individuals with a short or long immunity against re-infection. This means that it is natural to include the effects of immunity into the mathematical models in order to better represent the actual dynamics of epidemic spread and predict future outbreaks.  Immunity can be attained through targeted immunization, it can be naturally acquired after an individual has successfully recovered from an infection, and in some cases maternal antibodies can be transmitted to a newborn providing a certain level of immunity. In each case the immunity period will vary, as some diseases provide almost life-long immunity while others give only a very short-lived non-susceptibility. Quite often, the vaccine-induced immunity requires boosting after some period of time, as the vaccine effectiveness  wanes  due to absence of exposure to the disease. In the case of measles, for example,  vaccinated individuals are less immune than those with naturally acquired immunity \cite{MM,LvD}. Hepatitis B vaccination gives only $10$ to $15$ year immunity and after that  the boost is required in order for immunity to remain effective \cite{Lu}. In the case of serogroup C meningococcal disease, the immunity wanes with time and its efficacy strongly depends on the immunization programmes available \cite{WTP}. Recent outbreaks of mumps epidemic even amongst vaccinated population suggest that anti-mumps virus antibodies wane either due to a wild-type virus or because the population does not have sufficient exposures to the disease \cite{JOJD,Dayan}. Other diseases with waning immunity include varicella virus (although the cause of it is still being debated in the literature, some studies suggest that there is a need for a secondary vaccination dose \cite{BHHK,Arvin}), pertussis, for which immunity declines 6-12 years after the last episode of illness or booster dose \cite{GKVH}, and influenza which has a very short lived and strain-dependent immunity \cite{MMMMM}.

Delay differential equations have been successfully used to model varying infectious period in a range of SIR (susceptible-infected-recovered), SIS (susceptible-infected-susceptible) and SIRS (susceptible-infected-recovered-susceptible) epidemic models. Hethcote and van den Driessche have considered an SIS epidemic model with variable population size and constant time delay, which accounts for duration of infectiousness \cite{HvD1}. They found that an endemic equilibrium, when it exists, may undergo Hopf bifurcation and give rise to sustained periodic oscillations for certain parameter values. Beretta {\it et al.} \cite{Beretta} have studied global stability in an SIR epidemic model with distributed delay that describes the time it takes for an individual to lose infectiousness. They have used Lyapunov functionals to prove global stability of disease-free and endemic equilibria, provided the effective time delay is sufficiently small. Gao {\it et al.} \cite{Gao} have investigated the effects of pulse vaccination in an SIR model with distributed delay. They have shown that for sufficiently high vaccination rate, the disease-free periodic solution is globally attractive, which implies it is possible to completely eradicate the disease from the population. Another interesting integro-differential model was studied by Arino {\it et al.} \cite{Arino}, where the authors considered a vaccine, whose effectiveness wanes with time according to a general function. They have shown that the system can exhibit backward bifurcation and multi-stability for a range of parameter values, which has important implication for a design of optimal vaccination policies. A general discussion of time delay models in the context of epidemics can be found, for example, in Arino and van den Driessche \cite{AvD} (and references therein).

Some time ago we put forward a mathematical model of a disease with temporary immunity based on a system of delay differential equations \cite{KB}. Under quite general assumptions on transmission incidence, we found conditions for local and global stability of a disease-free and an endemic equilibrium when it exists. Several authors have since considered this model with particular choices of linear/non-linear incidence rate \cite{WY08,JW08}, and most recently, Brauer {\it et al.} have used it to study the spread of infection on a two-patch environment \cite{BV}. The main assumption was that there is a fixed duration of temporary immunity, after which recovered individuals return to the class of susceptibles. In this paper, we relax this assumption by considering a more realistic situation, in which immunity wanes with time. To model this, we introduce a delay differential equation model with distributed delay which takes into account varying immunity period. This means that after recovery an individual becomes susceptible again only after spending some time having acquired immunity from a disease. As it is important to predict the dynamics of an epidemic, we will prove local and global stability of the disease-free equilibrium. Biologically, this stability implies that although the disease may initially be present in the population, as time progresses it will eventually die out. Moreover, for some values of parameters it is possible to show that when an endemic equilibrium is feasible, it may be globally asymptotically stable. To get a better understanding of model dynamics, we will perform numerical bifurcation analysis with a Dirac $\delta$-function kernel using a DDE-BIFTOOL package \cite{ELS}. DDE-BIFTOOL is a numerical continuation tool that detects bifurcations and allows one to follow steady states and periodic solutions in the parameter space to get a general picture of system's behaviour depending on parameters and, in particular, on the time delay. For non-trivial kernels, i.e. weak and strong kernels, we will use traceDDE \cite{Breda} to find the boundary of Hopf bifurcation in terms of system parameters and an average time delay.

The paper is organised as follows. In Section 2 we introduce a model and prove that solutions of the model with positive initial conditions will remain positive for all time. Section 3 is devoted to the local and global stability analyses of disease-free and endemic equilibria. Numerical bifurcation analysis of the model is performed in Section 4 for several different choices of delay kernels. The paper concludes with summary and discussions.

\section{Derivation of the model and positivity of solutions}

In this section we will derive a delayed epidemic model with distributed delay. Recently, Kyrychko and Blyuss have introduced and studied a mathematical model for diseases with a constant immunity time and a nonlinear incidence rate in the form  \cite{KB}:

\begin{equation}\label{*}
\begin{array}{l}
\displaystyle \frac{dS(t)}{dt}=\mu-\phi f(I(t))S(t)-\mu
S(t)+\gamma I(t-\tau)e^{-\mu\tau},\\ \\
\displaystyle \frac{dI(t)}{dt}=\phi
f(I(t))S(t)-(\mu+\gamma)I(t),\\ \\
\displaystyle \frac{dR(t)}{dt}=\gamma I(t)-\gamma I(t-\tau)e^{-\mu\tau}-\mu R(t),
\end{array}
\end{equation}
where the total population is divided into three sub-groups, $S(t)$, $I(t)$ and $R(t)$ to denote susceptible, infected and recovered individuals,  respectively. The parameters are as follows, $\mu$ is a natural mortality rate, $\phi$ is a disease transmission rate, and $\gamma$ is a recovery rate.
Nonlinear incidence rate is represented by the function $f(I)$, and $\tau$ is a temporary immunity period, after which recovered individuals return into the class of susceptible. In this model, the immunity period is assumed to be constant and the same for all individuals. It is noteworthy that in the above model $S(t)+I(t)+R(t)=N(t)$, and $N(t)\to1$ as $t\to\infty$.

As it was described in the Introduction, the duration of immunity for different diseases may vary from a very short-lived to life-long immunity, and waning rates depend on the amount of exposure, boosting times, age etc. To account for this in our model, we assume that 
immunity period $\tau$ for different individuals varies from $0$ to $\infty$. We denote by $g(\xi)$ the probability density of taking time $\xi$ to lose acquired immunity, so that $g(\xi)d\xi$ is the probability of losing immunity somewhere between $\xi$ and $\xi+d\xi$ after acquiring it. If  one normalizes $g(\xi)$ as follows,
\begin{equation}\label{gfunc}
\int_{0}^{\infty} g(s)ds=1, \mbox{ and } g\geq 0,
\end{equation}
then the probability of having lost acquired immunity $s$ time units after acquiring it is $\displaystyle{\int_{0}^{s}g(\xi)d\xi}$, and the probability of still having immunity $s$ time units after acquiring it is $\displaystyle{1-\int_{0}^{s}g(\xi)d\xi=\int_{s}^{\infty}g(\xi)d\xi}$. The rate of recruitment into the recovered class $R$ at a time $t-s$ is $\gamma I(t-s)$. The probability of these individuals still being alive at time $t$ is $e^{-\mu s}$, and the probability of them still being immune is $\displaystyle{\int_{s}^{\infty}g(\xi)d\xi}$. Integrating over all previous times leads one to the expression
\begin{equation}\label{Req}
R(t)=\gamma\int_{0}^{\infty}I(t-s)e^{-\mu s}\int_{s}^{\infty}g(\xi)d\xi ds=\gamma\int_{-\infty}^{t}I(s)e^{-\mu(t-s)}\int_{t-s}^{\infty}g(\xi)d\xi ds.
\end{equation}
Differentiating this expression gives the following integro-differential equation for number of recovered individuals $R(t)$:
\[
\frac{dR}{dt}=\gamma I(t)-\gamma \int_{0}^{\infty}I(t-s)g(s)e^{-\mu s}ds-\mu R(t).
\]
With this derivation, the model to be investigated in this paper takes the form of a system of delay differential equations with distributed delay:
\begin{equation}\label{1}
\begin{array}{l}
\displaystyle\frac{dS(t)}{dt}=\mu-\phi f(I(t))S(t)-\mu
S(t)+\gamma
\int_{0}^{\infty}I(t-s)g(s)e^{-\mu s}ds,\\ \\
\displaystyle\frac{dI(t)}{dt}=\phi
f(I(t))S(t)-(\mu+\gamma)I(t),\\ \\
\displaystyle \frac{dR(t)}{dt}=\gamma I(t)-\gamma \int_{0}^{\infty}I(t-s)g(s)e^{-\mu s}ds-\mu R(t),
\end{array}
\end{equation}
where, as before, $\mu$ is a natural mortality rate, $\phi$ is a disease transmission rate, and $\gamma$ is a recovery rate.

It is easy to see that when $g(s)=\delta(s-\tau)$, where $\delta$ is the Dirac delta
function, the integral in (\ref{1}) becomes time-delayed term in (\ref{*}):
\[
\int_{0}^{\infty}I(t-s)\delta(s-\tau)e^{-\mu s}ds=e^{-\mu\tau}I(t-\tau).
\]
In order to simplify system (\ref{1}) we assume that transmission
of the disease can be adequately described as bilinear, and, consequently, the term
$f(I(t))S(t)$ becomes $I(t)S(t)$. With this assumption we arrive at the system
\begin{equation}\label{2}
\begin{array}{l}
\displaystyle\frac{dS(t)}{dt}=\mu-\phi I(t)S(t)-\mu
S(t)+\gamma
\int_{0}^{\infty}I(t-s)g(s)e^{-\mu s}ds,\\ \\
\displaystyle\frac{dI(t)}{dt}=\phi I(t)S(t)-(\mu+\gamma)I(t),\\ \\
\displaystyle \frac{dR(t)}{dt}=\gamma I(t)-\gamma \int_{0}^{\infty}I(t-s)g(s)e^{-\mu s}ds-\mu R(t),
\end{array}
\end{equation}
where the last equation may be omitted due to the fact that the first two equations do not depend on $R(t)$.

Since the model (\ref{2}) describes temporal dynamics of human population, it is important to show that the solutions of this system  do not become negative. We need to show that $S(t)>0$, $I(t)>0$, $R(t)>0$, i.e. the solutions of
system (\ref{2}) are positive for all $t\in(0;\infty)$. It is more convenient to start
by proving that $I(t)>0$ for all $t>0$. Let $I_{0}(s)\geq 0$,
$s\in(-\infty;0)$, and $I_{0}(0)>0$ be the initial data for
$I(t)$. We shall prove the positivity of $I(t)$ by contradiction.
Let $t_{1}>0$ be the first time when $I(t)S(t)=0$. Assuming that
$I(t_{1})=0$ implies that $S(t)\geq 0$, for all $t\in[0;t_{1}]$. Let
\[
A=\min_{0\leq t\leq t_{1}}\{\phi S(t)-\mu-\gamma\},
\]
then, for $t\in[0;t_{1}]$, $\displaystyle{dI/dt\geq AI(t)}$. Therefore, $I(t_{1})\geq I(0)e^{At_{1}}>0$. This is a
contradiction, and, hence, $I(t)>0$ for all $t>0$. Next, let
$S_{0}$ denote the initial data for $S(t)$, so that
$S(t)=S_{0}(t)$ for all $t\in (-\infty;0)$. Let $S_{0}$ be
continuous and satisfy $S_{0}(t)\geq 0$ for all $t\in(-\infty;0)$,
and $S_{0}(0)>0$. By contradiction, we prove that $S(t)>0$ for all
$t>0$. Assume that there exists a first time $t_{0}>0$ such that
$S(t_{0})=0$. This indicates that $S(t)>0$ for $t\in[0;t_{0})$ and
\[
\frac{dS(t_{0})}{dt}=\mu+\gamma\int_{0}^{\infty}\underbrace{I(t_{0}-s)}_{>0,\forall
t_{0}\in[0;\infty) }g(s)e^{-\mu s}ds>0.
\]
Therefore, $\displaystyle\frac{dS(t_{0})}{dt}>0$. This is a contradiction
to our assumption, since it implies $S(t)$ must be negative for
$t$ just before $t_{0}$, which contradicts the choice of $t_{0}$. 
The positivity of $R(t)$ for $t\geq 0$ follows from the integral representation (\ref{Req}) and positivity of $I(t)$.
Therefore, we have proved that the solutions of system (\ref{2}) with positive initial conditions remain positive for all times.

\section{Equilibria and their stability}

\subsection{Local stability of the disease-free and endemic steady states}

Omitting the last equation in system (\ref{2}), equilibria
$(\bar{S},\bar{I})$ are determined by setting
$\dot{S}(t)=\dot{I}(t)=0$.
There are two steady states, namely, a disease-free steady state $E_{0}=(1;0)$ and an endemic steady state
$\widetilde{E}=(\widetilde{S},\widetilde{I})$, where
\[
\widetilde{S}=\frac{\gamma+\mu}{\phi}\mbox{ and }
\widetilde{I}=\frac{\mu(\gamma+\mu)-\mu\phi}{\phi\left(-\gamma-\mu+\gamma\displaystyle{\int_{0}^{\infty}g(s)e^{-\mu
s}ds}\right)}.
\]
Using the properties of function $g(s)$ from (\ref{gfunc}), one concludes that while the equilibrium $\widetilde{E}$ exists for arbitrary values of parameters, it is biologically relevant if and only if
\[
\gamma+\mu<\phi.
\]

It is important to analyse the stability of these equilibria, as it will indicate whether the disease will die out eventually, or it will persist for all time. The linearisation of the system (\ref{2}) without the last equation near the steady state $E_{0}=(1,0)$ gives
\begin{equation}\label{3}
\begin{array}{l}
\displaystyle\frac{d\widehat{S}(t)}{dt}=\phi \widehat{I}-\mu
\widehat{S}(t)+\gamma\int_{0}^{\infty}\widehat{I}(t-s)g(s)e^{-\mu s}ds,\\\\
\displaystyle\frac{d\widehat{I}(t)}{dt}=\phi
\widehat{I}(t)-(\mu+\gamma)\widehat{I}(t).
\end{array}
\end{equation}
The characteristic equation for (\ref{3}) has the form
\[
(\lambda+\mu)(\lambda-\phi+\mu+\gamma)=0,
\]
with the corresponding eigenvalues
\begin{equation}\label{4}
\lambda_{1}=-\mu;\mbox{ }\lambda_{2}=\phi-\mu-\gamma.
\end{equation}
We define the basic reproduction number as
\[
\mathcal{R}_{0}=\frac{\phi}{\mu+\gamma}.
\]
The basic reproduction number identifies the number of secondary infections from the infected source and plays a crucial role in understanding the development of epidemics.
From (\ref{4}) it follows that when $\mathcal{R}_{0}<1$, the
disease-free equilibrium $E_{0}=(1,0)$ is the only steady state of the model, and it is locally
asymptotically stable. 

When $\mathcal{R}_{0}>1$, there exists a
non-trivial equilibrium $\widetilde{E}=(\widetilde{S},\widetilde{I})$. In order to analyse the stability of this equilibrium we use a Lyapunov functional technique. Linearising the
system (\ref{2}) without the last equation about $\widetilde{E}$ by setting $S(t)=\widetilde{S}+p(t)$
and $I(t)=\widetilde{I}+q(t)$ gives
\begin{equation}\label{6}
\begin{array}{l}
\displaystyle\frac{dp(t)}{dt}=(-\phi \widetilde{I}-\mu)
p(t)-\phi\widetilde{S}q(t)+\gamma\int_{0}^{\infty}q(t-s)g(s)e^{-\mu s}ds,\\ \\
\displaystyle\frac{dq(t)}{dt}=\phi \widetilde{I}p(t),
\end{array}
\end{equation}
where $\displaystyle\widetilde{S}=\frac{\gamma+\mu}{\phi}$.
Introduce the following functional:
\[
V=\frac{1}{2}q^{2}(t)+\frac{w}{2}(p(t)+q(t))^{2}+w\gamma\int_{0}^{\infty}
g(s)e^{-2\mu s}\int_{t-s}^{t}q^{2}(\nu)d\nu ds,
\]
where $w$ is a real positive constant, and $V(t)\geq 0$.
Differentiating $V$ along the solution of system (\ref{6}) and substituting $\dot{q}$ and $\dot{p}$ we
arrive at
\begin{eqnarray}\label{7}
\dot{V}=&&\phi\widetilde{I}pq-w\mu p^{2}-w(\mu+\gamma)pq+w\gamma
p\int_{0}^{\infty}q(t-s)g(s)e^{-\mu s}ds\nonumber\\
&-&w\mu pq-w(\mu+\gamma)q^{2}
+wq\gamma\int_{0}^{\infty}q(t-s)g(s)e^{-\mu s
}ds\nonumber\\
&+&w\gamma q^{2}\int_{0}^{\infty}g(s)e^{-2\mu s}ds
-w\gamma\int_{0}^{\infty}g(s)e^{-2\mu s}q^{2}(t-s)ds.
\end{eqnarray}
Estimating the fourth and seventh terms using Cauchy-Schwarz and then H\"older inequalities, 
one can rewrite (\ref{7}) as
\[
\dot{V}\leq pq\left[\phi\widetilde{I}-w\mu-w(\mu+\gamma)\right]-w\mu p^{2}-w(\gamma+\mu)q^{2}+\frac{w\gamma}{2}q^{2}+\frac{w\gamma}{2}p^{2}+w\gamma q^{2}\underbrace{\int_{0}^{\infty}g(s)e^{-2\mu s}ds}_{\leq 1}.
\]
Since $\widetilde{I}>0$, we can choose a positive constant $w$ as $w=\phi\widetilde{I}/(2\mu+\gamma)$.
Therefore, the above inequality simplifies to
\[
\dot{V}\leq-\left[\mu-\frac{\gamma}{2}\right]w(p^{2}+q^{2}).
\]
As the expression in the right-hand side is negative-definite, provided $\mu>\gamma/2$, we have proved the following theorem:\\

\noindent {\bf Theorem 1.} {\it Whenever $\phi>\mu+\gamma$, the endemic equilibrium
$\widetilde{E}=(\widetilde{S},\widetilde{I})$ is feasible, and provided $\mu>\gamma/2$, it is locally asymptotically stable}.

\subsection{Global stability of the disease-free steady state}

It has already been shown that when $\phi<\mu+\gamma$, the trivial equilibrium $E_{0}$ of the system (\ref{2}) is locally stable. In fact, in this case we have the following result.\\

\noindent {\bf Theorem 2.} {\it If $\phi<\mu+\gamma$, the disease-free equilibrium $E_{0}=(1,0)$ is globally asymptotically stable}.\\

\noindent{\it Proof}. To prove global stability of the disease-free steady state, we first define $N(t)=S(t)+I(t)+R(t)$. Adding all three equations in (\ref{2}) gives
\[
\frac{dN}{dt}=\mu-\mu N(t),
\]
which implies $\lim_{t\to\infty}N(t)=1$. Under the assumption $\phi<\mu+\gamma$, we may choose $\epsilon>0$ sufficiently small, such that $\phi(1+\epsilon)<\mu+\gamma$. Since $N(t)\to 1$ as $t\to\infty$, for sufficiently large $t$ we have $S(t)+I(t)+R(t)=N(t)\leq 1+\epsilon$. Then the equation for $I(t)$ becomes
\begin{eqnarray*}
\frac{dI}{dt}&\leq& \phi I(t)\left[1+\epsilon-I(t)-R(t)\right]-(\mu+\gamma)I(t)\\
&\leq& \phi I(t)[1+\epsilon-I(t)]-(\mu+\gamma)I(t)\\
&=&I(t)\left[\phi(1+\epsilon)-\phi I(t)-(\mu+\gamma)\right].
\end{eqnarray*}
From a simple comparison argument and using the fact that $\phi(1+\epsilon)<\mu+\gamma$, it follows that $I(t)\to 0$ as $t\to\infty$. From equation (\ref{Req}) for $R(t)$ it follows that $R(t)\to 0$ as $t\to\infty$. Since $N(t)\to 1$, it follows that $S(t)\to 1$ as $t\to\infty$, which implies global asymptotic stability of the trivial steady state.\hfill$\blacksquare$

\subsection{Global stability of endemic steady state}

As we have already established, when $\phi>\mu+\gamma$ the disease-free steady state is unstable, and there is a feasible endemic equilibrium $\widetilde{E}=(\widetilde{S},\widetilde{I},\widetilde{R})$, which according to Theorem 1 is locally asymptotically stable. To prove global stability of this steady state, we will employ Lyapunov functional approach. Let us introduce new variables $u_{1}(t)=S(t)-\widetilde{S}$, $u_{2}(t)=I(t)-\widetilde{I}$, $u_{3}(t)=R(t)-\widetilde{R}$. With this change of variables, system (\ref{2}) can be written in the form
\begin{equation}\label{end_sys}
\begin{array}{l}
\displaystyle{\frac{du_{1}}{dt}=-\mu u_{1}-\phi Su_{2}-\phi u_{1}\widetilde{I}+\gamma\int_{0}^{\infty}u_{2}(t-s)g(s)e^{-\mu s}ds,}\\\\
\displaystyle{\frac{du_{2}}{dt}=\phi Su_{2}+\phi u_{1}\widetilde{I}-(\mu+\gamma)u_{2},}\\\\
\displaystyle{\frac{du_{3}}{dt}=\gamma u_{2}-\gamma\int_{0}^{\infty}u_{2}(t-s)g(s)e^{-\mu s}ds-\mu u_{3}.}
\end{array}
\end{equation}
Global stability of the trivial equilibrium of system (\ref{end_sys}) implies global stability of an endemic equilibrium $\widetilde{E}$ of the original system (\ref{2}). Introduce the following functional
\begin{equation}
V(u)=\frac{1}{2}w(u_{1}+u_{2})^{2}+\frac{1}{2}\left(u_{2}^{2}+u_{3}^{2}\right),
\end{equation}
where $w$ is a positive constant to be determined later in the calculations.
Differentiating $V(u)$ and using (\ref{end_sys}) gives
\begin{eqnarray*}
\dot{V}(u)=&-&\mu w u_{1}^{2}-(w+1)(\mu+\gamma)u_{2}^{2}-\mu u_{3}^{2}+\phi S u_{2}^{2}\\
&+&u_{1}u_{2}[-(\mu+\gamma)w-\mu w+\phi\widetilde{I}]+wu_{1}\gamma\int_{0}^{\infty}u_{2}(t-s)g(s)e^{-\mu s}ds\\
&+&w\gamma u_{2}\int_{0}^{\infty}u_{2}(t-s)g(s)e^{-\mu s}ds+\gamma u_{2}u_{3}-\gamma u_{3}\int_{0}^{\infty}u_{2}(t-s)g(s)e^{-\mu s}ds.
\end{eqnarray*}
It can be easily shown that for non-negative initial conditions $S(0)=S_{0}>0$, $I(s)=I_{0}(s)\geq 0$ for all $s\in(-\infty,0)$ with $I(0)=I_{0}>0$ and $R(0)=R_{0}>0$, one has $S(t)\leq \max\{1,S_{0}+I_{0}+R_{0}\}=M$ for all $t>0$ \cite{KB}. Choosing $w=\phi\widetilde{I}/(2\mu+\gamma)$ in the above equation leads to
\begin{eqnarray}\label{lyap1}
\dot{V}&\leq&-\mu w u_{1}^{2}-[(w+1)(\mu+\gamma)-\phi M]u_{2}^{2}-\mu u_{3}^{2}+wu_{1}\gamma\int_{0}^{\infty}u_{2}(t-s)g(s)e^{-\mu s}ds\nonumber\\
&+&w\gamma u_{2}\int_{0}^{\infty}u_{2}(t-s)g(s)e^{-\mu s}ds+\gamma u_{2}u_{3}-\gamma u_{3}\int_{0}^{\infty}u_{2}(t-s)g(s)e^{-\mu s}ds.
\end{eqnarray}
The fourth term can be estimated with the help of Cauchy-Schwartz and H\"older inequalities as follows
\[
wu_{1}\gamma\int_{0}^{\infty}u_{2}(t-s)g(s)e^{-\mu s}ds\leq \frac{w \gamma}{2}u_{1}^{2}+\frac{w\gamma}{2}\int_{0}^{\infty}u_{2}^{2}(t-s)g(s)e^{-2\mu s}ds,
\]
and similar estimates can be made for other $u_{i}u_{j}$, $i\neq j$ terms in (\ref{lyap1}).
This finally gives
\begin{eqnarray}\label{lyap2}
\dot{V}(u)\leq &-&w\left(\mu-\frac{\gamma}{2}\right)u_{1}^{2}-\left[\left(\mu+\frac{\gamma}{2}\right)(w+1)-\phi M\right]u_{2}^{2}\nonumber\\
&-&(\mu-\gamma)u_{3}^{2}+\gamma \left(w+\frac{1}{2}\right)\int_{0}^{\infty}u_{2}^{2}(t-s)g(s)e^{-2\mu s}ds.
\end{eqnarray}
Let now the Lyapunov functional for the system (\ref{end_sys}) be
\[
U(u)=V(u)+\gamma\left(w+\frac{1}{2}\right)\int_{0}^{\infty}g(s)e^{-2\mu s}\int_{t-s}^{t}u_{2}^{2}(\nu)d\nu ds.
\]
Differentiating $U$ and using (\ref{lyap2}), we obtain
\begin{eqnarray*}
\dot{U}(u)&=&\dot{V}(u)+\gamma\left(w+\frac{1}{2}\right)u_{2}^{2}(t)\int_{0}^{\infty}g(s)e^{-2\mu s}ds-\gamma\left(w+\frac{1}{2}\right)\int_{0}^{\infty}u_{2}^{2}(t-s)g(s)e^{-2\mu s}ds\\
&\leq& -w\left(\mu-\frac{\gamma}{2}\right)u_{1}^{2}-
\left[\left(\mu+\frac{\gamma}{2}\right)(w+1)-\phi M\right]u_{2}^{2}-(\mu-\gamma)u_{3}^{2}\\
&+&\gamma \left(w+\frac{1}{2}\right)u_{2}^{2}(t)\underbrace{\int_{0}^{\infty}g(s)e^{-2\mu s}ds}_{\leq 1}.
\end{eqnarray*}
Finally, the Lyapunov functional can be estimated as follows
\[
\dot{U}(u)\leq -w\left(\mu-\frac{\gamma}{2}\right)u_{1}^{2}-
\left[\mu(w+1)-\frac{w \gamma}{2}-\phi M\right]u_{2}^{2}-(\mu-\gamma)u_{3}^{2},
\]
which is negative-definite if $\mu>\gamma$ and $\mu(w+1)-w \gamma/2-\phi M>0$. From Lyapunov-LaSalle type theorem (Theorem 2.5.3 of Kuang \cite{Kuang}), it then follows that $\lim_{t\to\infty}u_{k}(t))=0$, $k=1,2,3$.
Thus, we have proved the following result.\\

\noindent{\bf Theorem 3.} {\it Let the initial conditions for system (\ref{2}) be $S(0)=S_{0}>0$, $I(s)=I_{0}(s)\geq 0$ for all $s\in(-\infty,0)$ with $I(0)=I_{0}>0$ and $R(0)=R_{0}>0$. If 
\[
\mu>\gamma,\hspace{0.5cm}\phi>\mu+\gamma,\hspace{0.5cm}\mu(w+1)-w \gamma/2-\phi M>0,
\]
where $w=\phi\widetilde{I}/(2\mu+\gamma)$ and $M=\max\{1,S_{0}+I_{0}+R_{0}\}$, then the endemic steady state $\widetilde{E}$ of system (\ref{2}) is globally asymptotically stable.}\\

Since biologically reasonable initial conditions for the system (\ref{2}) satisfy $S_{0}+I_{0}+R_{0}=1$, the conditions of Theorem 3 depend {\it only} on system parameters, and the Theorem 3 then holds for all initial conditions.

\begin{figure*}
\hspace{-1cm}\includegraphics[width=18cm]{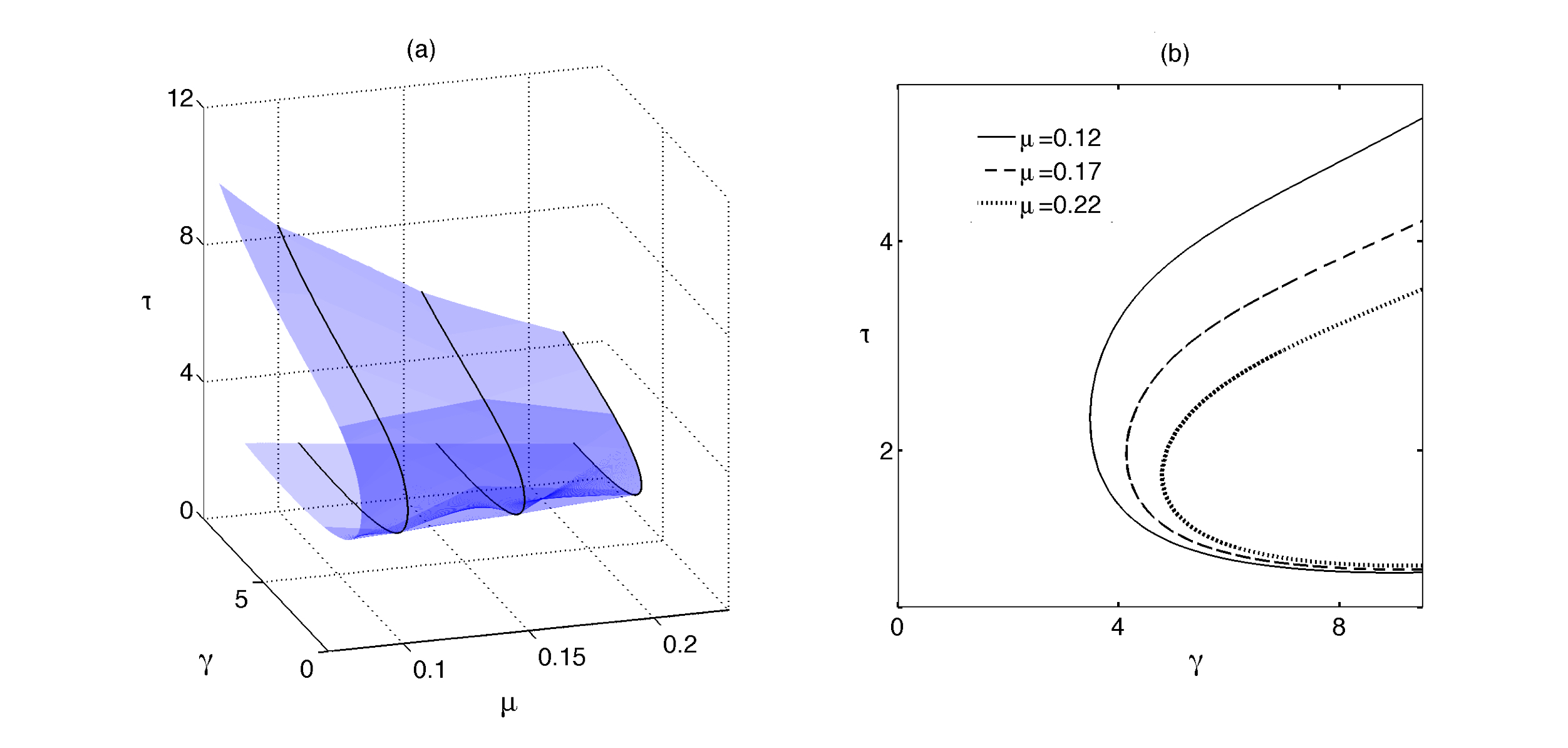}
\caption{(a) Boundary of Hopf bifurcation of the endemic steady state in terms of $\tau$, $\gamma$ and $\mu$. (b) Two-parameter continuation of Hopf boundary for different values of natural death rate $\mu$. In both pictures $\phi=10$.}\label{gt}
\end{figure*}

\section{Bifurcations and continuation}

In this section we perform numerical bifurcation analysis of an epidemic model (\ref{2}).  The main purpose is to get an insight into how the dynamics of the system changes depending on the system parameters, and in particular, on the length of the immunity period. We shall proceed by finding bifurcation points of the endemic steady state in terms of time delay $\tau$, and then continue the bifurcating solutions by changing either the transmission rate $\phi$, the death rate $\mu$, or the recovery rate $\gamma$. Continuation analysis is an important part in the study of epidemic models, where parameter values carry a lot of uncertainty due the fact that some of them are difficult to obtain from experimental data, and some of them vary significantly in the population. Compared to straightforward numerical simulations, bifurcation analysis provides a much more complete picture of the underlying dynamical behaviour for a whole range of parameter values. A powerful continuation tool for delay differential equations is a MATLAB based tool called DDE-BIFTOOL \cite{ELS}. It has been successfully used for detecting and analysing bifurcations in models of laser dynamics \cite{Kraus}, car-following models \cite{OKW}, neural networks of coupled cells \cite{BC} and many other applications. Currently, DDE-BIFTOOL does not have capabilities for studying systems with distributed delay, so we will use traceDDE package \cite{Breda} instead.

\subsection{Fixed immunity period}

First of all, we consider the case when the integral kernel is the Dirac $\delta$-function $g(s)=\delta(s-\tau)$, which corresponds to a fixed period of temporary immunity $\tau$. Two particular choices of the transmission function $f(I)$ have been analysed, a linear transmission rate $f(I)=I$ and a saturated transmission rate in the form $f(I)=I/(1+I)$. The results of numerical bifurcation analysis are qualitatively similar for both of these functions, hence we only present here the computations for $f(I)=I$ to complement the analytical results from earlier sections. In all computations, the time delay $\tau$ plays a role of the main bifurcation parameter. After finding an endemic steady state, we then detect its bifurcations depending on the values of $\tau$ and other system parameters. 

\begin{figure*}
\hspace{-1cm}\includegraphics[width=18cm]{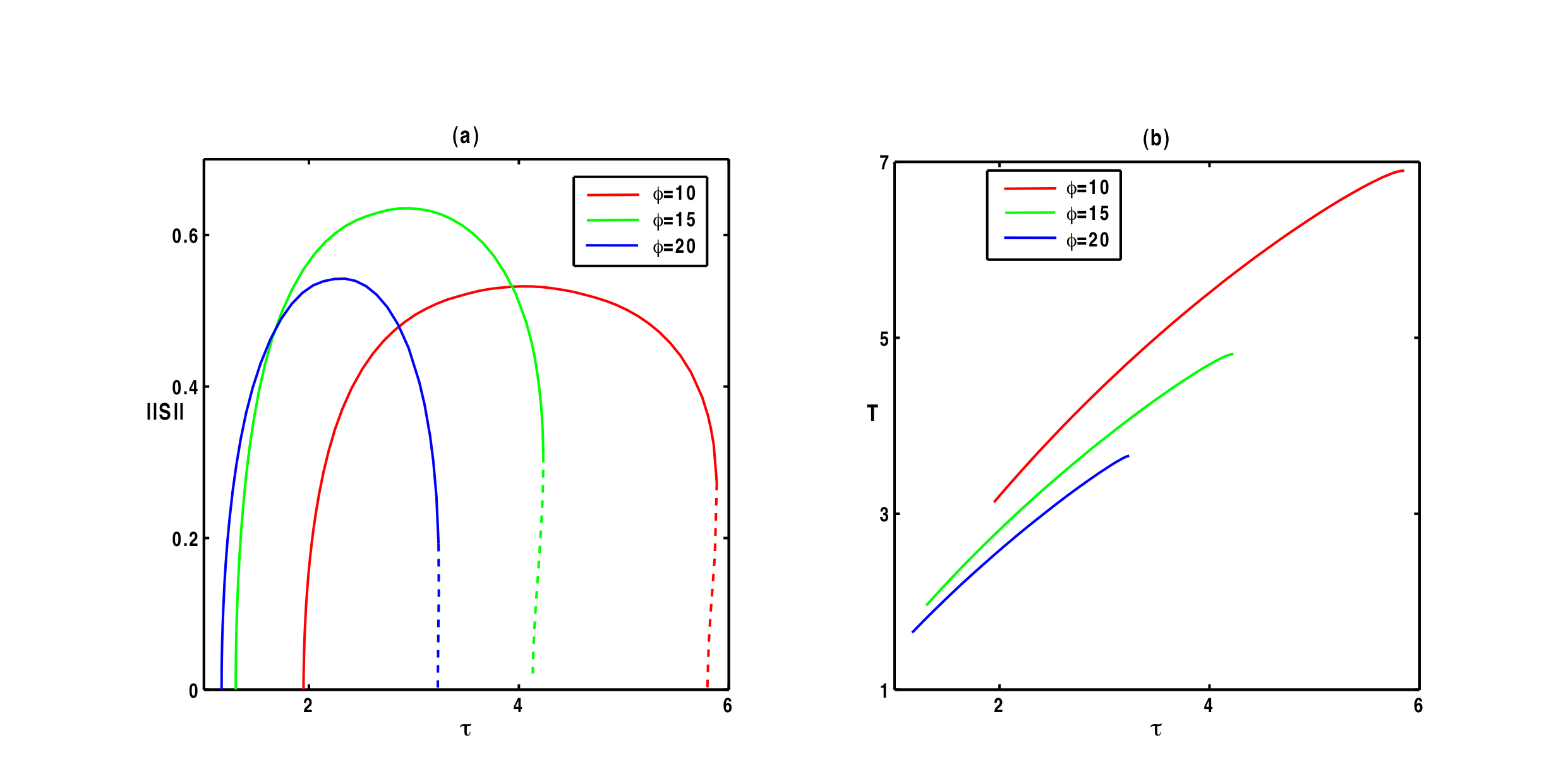}
\caption{(a) Amplitude of periodic solutions as a function of time delay $\tau$. Solid lines denote stable periodic orbits, dashed lines denote unstable periodic orbits. (b) Period $T$ of the periodic orbits depending on time delay. Parameter values are $\mu=0.15$ and $\gamma=5$.}\label{ampl}
\end{figure*}

Figure~\ref{gt} shows the boundary of Hopf bifurcation of the endemic steady state $\widetilde{E}$ in the space of parameters. To find this Hopf boundary, we fix the natural death rate $\mu$ and perform a two-parameter continuation in terms of a temporary immunity time delay $\tau$ and a recovery rate $\gamma$. It is noteworthy that the smaller the death rate $\mu$, the larger is the region covered by Hopf boundary in $\gamma$-$\tau$ plane. Furthermore, for each fixed $\gamma$, the periodic orbit corresponding to the smaller value of $\tau$ is stable, while its counterpart for the larger value of $\tau$ is unstable. Below the Hopf boundary, the endemic steady state $\widetilde{E}$ is stable.

Next, we take as a starting point a periodic orbit of very small amplitude close to the Hopf boundary, and continue this orbit in time delay $\tau$.  The results of this continuation are shown in Fig.~\ref{ampl}(a). They indicate that for sufficiently small $\tau$, the corresponding periodic orbit is stable (solid lines), but then it goes through a fold, and for sufficiently high values of the time delay $\tau$ there is narrow region of a co-existence of stable and unstable periodic orbits (dashed lines). The corresponding plot of the periods of those periodic orbits depending on the time delay is shown in Fig.~\ref{ampl}(b). This figure indicates that the period increases with the time delay, and decreases with the increase of transmission rate $\phi$. One can explain this observation by the fact that for a higher transmission rate $\phi$, it takes a shorter period of time for individuals to go through a disease cycle and return to the class of susceptibles, hence reducing the duration of a periodic cycle.

\begin{figure*}
\hspace{-1cm}\includegraphics[width=18cm]{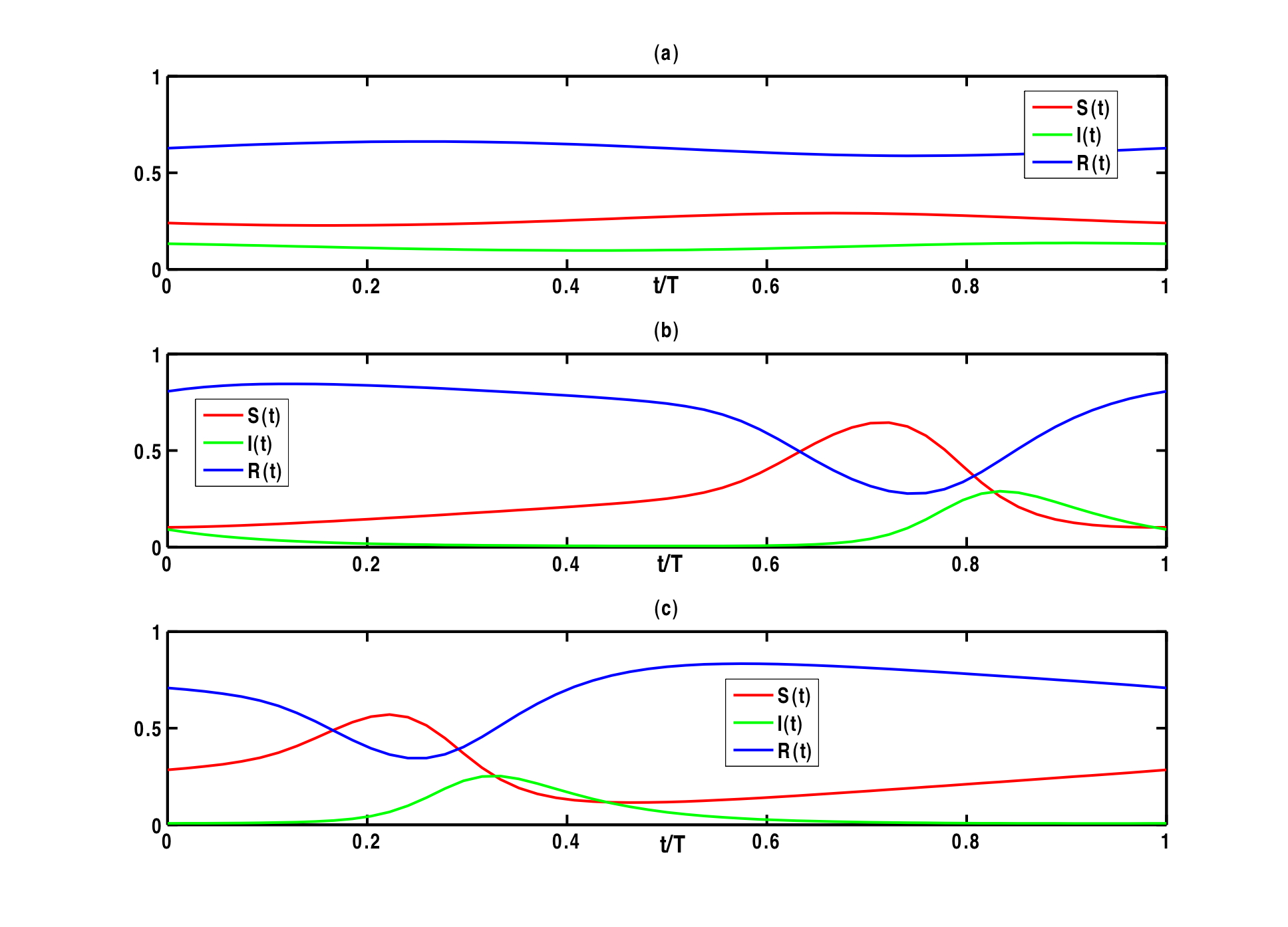}
\caption{Profiles of periodic solutions for $\mu=0.15$, $\gamma=5$ and $\phi=15$. The duration of temporary immunity period is (a) $\tau=1.1739$, (b) $\tau=2.3431$, and (c) $\tau=2.9367$.}\label{persol}
\end{figure*}

The temporary profiles of periodic solutions for different time delays are shown in Fig.~\ref{persol}, where all three cases represent stable periodic orbits. For $\tau$ just above the boundary of Hopf bifurcation, the solution shows little variation in either of the variables, but it becomes more pronounced for higher $\tau$. As is natural to expect, the peak of infectives slightly lags behind the peak of susceptibles, and one can also note a certain asymmetry in the profile of the solutions.

\subsection{Gamma distribution}

A more realistic representation of temporary immunity is achieved when the integral kernel is given by some non-trivial function $g(s)$. One possible choice is the so-called Gamma distribution delay kernel \cite{Ruan}
\[
g(s)=\frac{\alpha^{n}s^{n-1}e^{-\alpha s}}{(n-1)!},\hspace{0.5cm}n=1,2,...,
\]
where $\alpha>0$ is a constant. In this case, one can introduce an {\it average delay} as
\[
\bar{\tau}=\int_{0}^{\infty}sg(s)ds=n/\alpha.
\]
For simulations, we concentrate on two particular cases: $n=1$ and $n=2$, which correspond to a weak delay kernel and a strong delay kernel, respectively. In the case of weak kernel $g_{w}(s)=\alpha e^{-\alpha s}$, the maximum influx of recovered individuals into the class of susceptibles comes from individuals who are currently recovering, while past recoveries have exponentially decreasing contributions. On the other hand, for the strong kernel $g_{s}(s)=\alpha^{2}se^{-\alpha s}$, the maximum influx of recovered into susceptibles at any time $t$  comes from those who recovered at $t-\bar{\tau}$, where $\bar{\tau}$ is the average immunity period.

\begin{figure*}
\hspace{-0.5cm}\includegraphics[width=18cm]{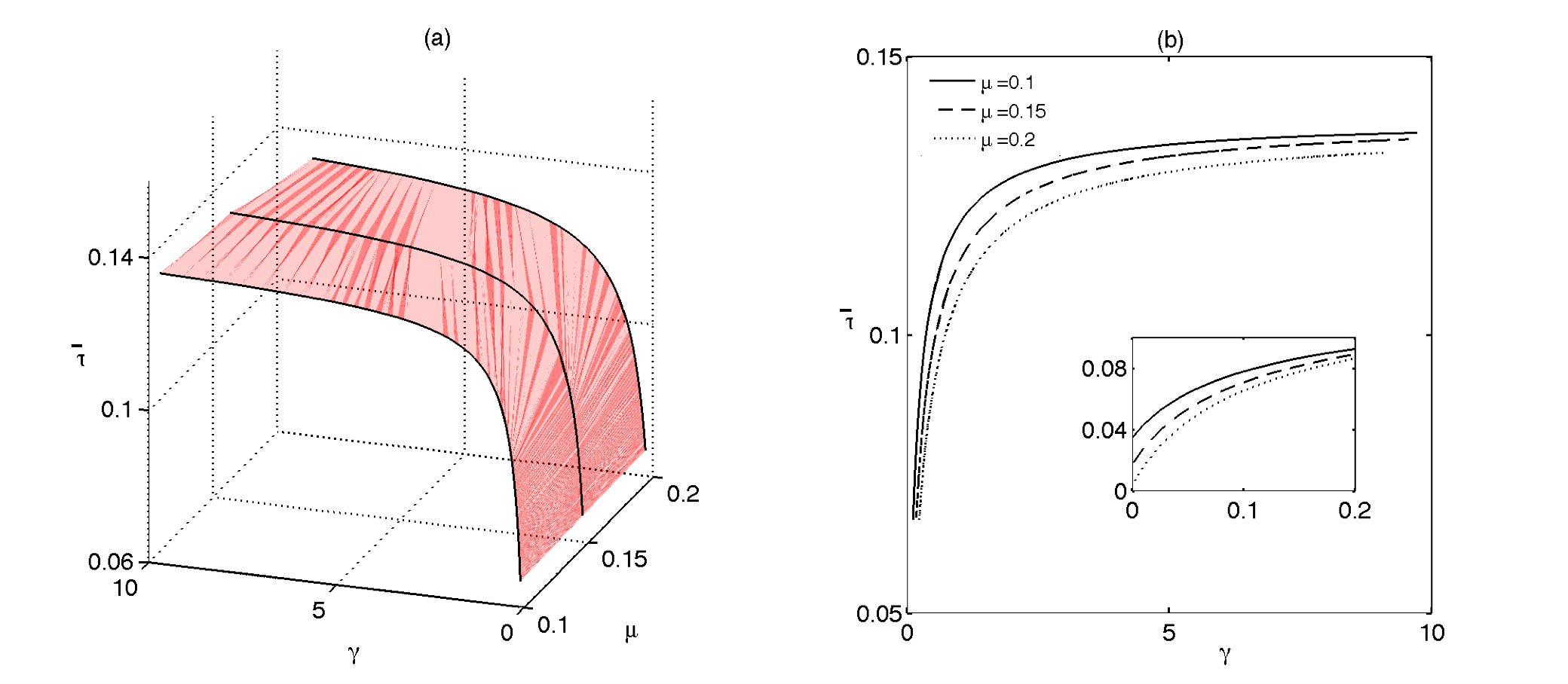}
\caption{(a) Boundary of the Hopf bifurcation of the endemic steady state with a weak kernel $g_{w}(s)=\alpha e^{-\alpha s}$ in terms of $\bar{\tau}$, $\gamma$ and $\mu$. The average immunity time is $\bar{\tau}=1/\alpha$. (b) Two-parameter continuation of Hopf boundary for different values of $\mu$. In both pictures $\phi=10$.}\label{hopf_weak}
\end{figure*}

For numerical bifurcation analysis of system (\ref{2}) with weak and strong kernels, we use a Matlab package traceDDE, which is based on pseudo-spectral differentiation and allows one to find characteristic roots and stability charts for linear autonomous systems of delay differential equations \cite{Breda}. In all simulations we replaced the upper limit of the integral by a large positive constant $L$, such that $e^{-\mu L}\ll 1$.

Figure~\ref{hopf_weak} depicts the boundary of Hopf bifurcation  for weak kernel in the space of average time delay $\bar{\tau}$, recovery rate $\gamma$ and natural mortality $\mu$. Endemic state is stable below the boundary and unstable above it. While the dependence on $\mu$ is weak, there is a noticeable variation in the critical $\bar{\tau}$ as a function of $\gamma$: the higher is the recovery rate $\gamma$, the higher is the average immunity period $\bar{\tau}$, at which the endemic steady state loses stability. We note that unlike the case of a Dirac $\delta$-function kernel, the dependence of $\bar{\tau}$ on $\gamma$ at the Hopf boundary is monotonic.

\begin{figure*}
\hspace{-0.5cm}\includegraphics[width=17.5cm]{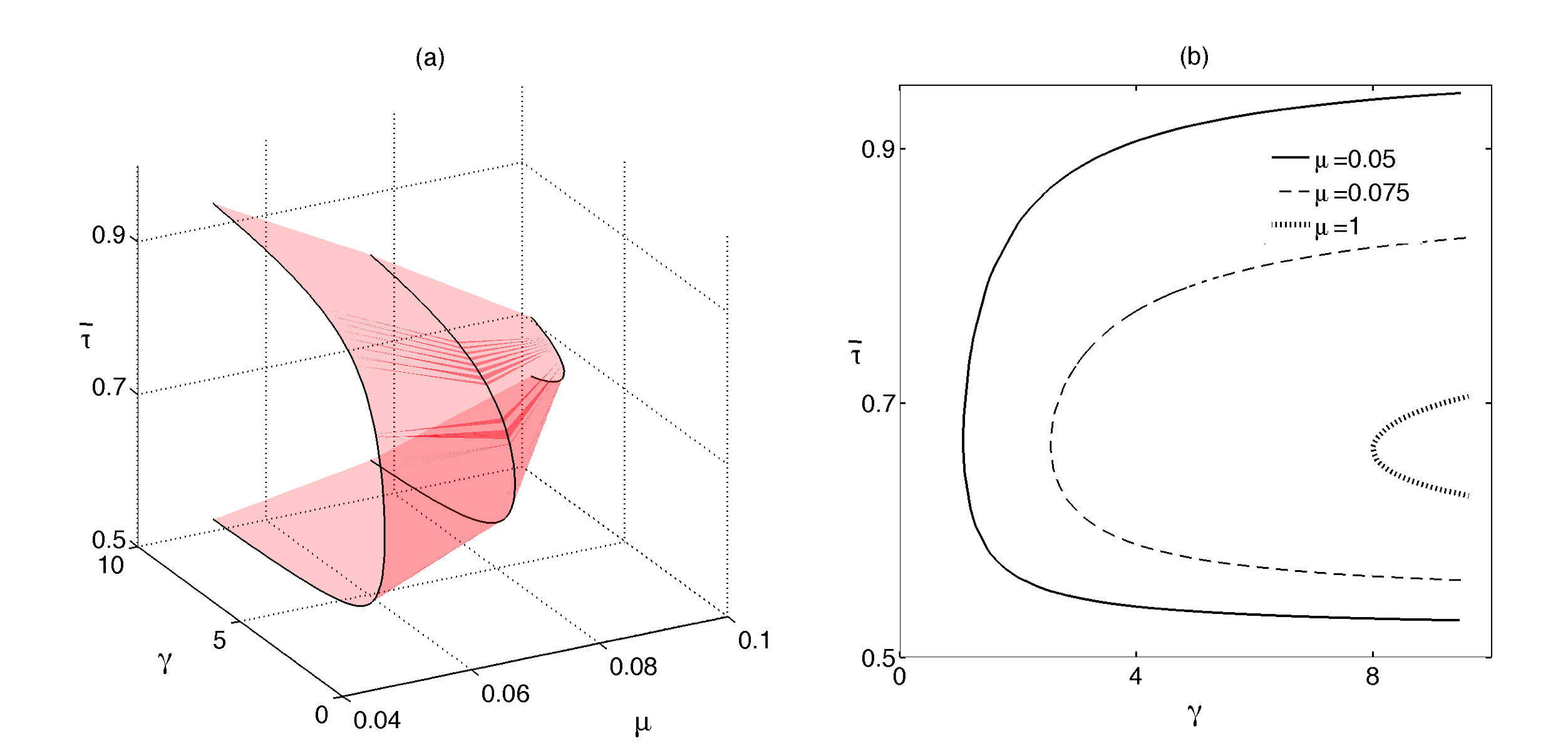}
\caption{(a) Boundary of the Hopf bifurcation of the endemic steady state with a strong kernel $g_{s}(s)=\alpha^{2}se^{-\alpha s}$ in terms of $\bar{\tau}$, $\gamma$ and $\mu$. The average immunity time is $\bar{\tau}=2/\alpha$. (b) Two-parameter continuation of Hopf boundary for different values of $\mu$. In both pictures $\phi=10$.}\label{hopf_strong}
\end{figure*}

In Fig.~\ref{hopf_strong} we show Hopf boundary in the case of a strong kernel $g_{s}(s)$. This figure bears some similarity to Figure~\ref{gt} due to the fact that, as we have already mentioned, the largest contribution in the strong kernel comes from the individuals who recovered time $\bar{\tau}$ ago, which is reminiscent of a $\delta$-function kernel $g(s)=\delta(s-\tau)$. The endemic steady state is stable outside the boundary and unstable inside. It is noteworthy that as the natural mortality $\mu$ increases, the instability region shrinks, and instability occurs for higher values of recovery rate $\gamma$.

\section{Conclusions}

In this paper we have derived a delay differential equations model for the dynamics of infectious diseases with varying temporary immunity. This model allows temporary immunity to wane with time, and is, therefore, a more realistic analogue of previous models which assumed fixed duration of immunity period. If transmission coefficient is not too high, the model admits a single disease-free equilibrium, which is globally asymptotically stable. For higher values of a disease transmission coefficient, there is a feasible endemic equilibrium, whose local and global stability has been proven for certain parameter values using a Lyapunov functional approach.

In order to obtain a better insight into the dynamics of the system in different parameter regimes, we have performed numerical bifurcation analysis for several particular choices of the integral kernel, including fixed temporary immunity, weak and strong kernels. The results of these simulations suggest that endemic equilibrium may lose its stability via a supercritical Hopf bifurcation, thus giving rise to stable periodic solutions. Biologically this means that when the temporary immunity period is within a certain range, there will be periodic outbreaks of epidemic, and the disease will not be eradicated from the population. In the case of delay kernel being a Dirac $\delta$-function, as the duration of temporary immunity $\tau$ increases, these periodic solutions lose stability at a fold bifurcation. The period of periodic orbits increases monotonically with $\tau$ and decreases with the increasing disease transmission coefficient. For a weak kernel, the average time delay corresponding to the Hopf boundary increases monotonically with the recovery rate $\gamma$. The case of strong kernel is somewhat similar to the case of Dirac $\delta$-function in that the endemic steady state is unstable for a range of average immunity periods, and the parameter region of instability shrinks with the increase of mortality rate $\mu$.

The model developed in this paper relates to some earlier work on modelling diseases with temporary immunity. For example, Cooke and van den Driessche have considered a model with constant temporary immunity and latency \cite{CvD}. They have shown that similarly to our model, it is possible to have periodic solutions for some parameter ranges. It is noteworthy that these periodic solution occur due to the time delay representing temporary immunity, but there is no such effect achieved by considering time delay in latency alone. In comparison, Gomes {\it et al.} have analysed several ODE models with temporary and partial immunity, as well as vaccination, where immunity wanes at a constant rate \cite{Gomes}. They have shown that when a basic reproduction number exceeds unity, the solutions either decay linearly to an endemic steady state, or approach it in an oscillatory manner. In their findings, immunity waning rate plays an important role in the time scale for oscillatory behaviour. While our model also highlights the importance of temporary immunity, in contrast to the above model it is also able to sustain periodic solutions describing regular epidemic outbreaks. The main feature is that temporary immunity leads to a possible destabilization of endemic steady state, and an interesting open question is what effects would vaccination have on the dynamics of an epidemic in such situation.

\section*{Acknowlegment} Y.K. was partially supported by the EPSRC (Grant EP/E045073/1). The authors would like to thank the anonymous referees for their helpful comments and suggestions.

\end{document}